\documentclass[twoside]{article}
\usepackage{amsmath}
\usepackage{amssymb}
\usepackage{cite}
\usepackage{graphicx}

\input ijmpblat
\newcommand{\nn}{\nonumber}

\newcommand{\kb}{k_{_{\mathrm{B}}}}
\newcommand{\eps}{\varepsilon}
\newcommand{\bp}{\mathbf{p}}
\newcommand{\la}{\left<}
\newcommand{\ra}{\right>}

\newcommand{\Tc}{\ensuremath{T_{\mathrm{c}}}}
\newcommand{\jour}[4]{{\nineit #1}\ {\ninebf #2},\ #3\ (#4)}
\newcommand{\Fref}[1]{Fig.~\ref{#1}}
\newcommand{\Eref}[1]{(\ref{#1})}
\newcommand{\Rref}[1]{Ref.~\citen{#1}}
\newcommand{\PRL}{Phys. Rev. Lett.}
\newcommand{\PR}{Phys. Rev.}

\begin{document}
\runninghead{T. Mishonov,  S.-L. Drechsler \& E. Penev}{Influence of
            the van Hove singularity on the specific heat jump in BCS
            superconductors}
%
\thispagestyle{empty}
\setcounter{page}{1}
%
%
\vspace*{0.88truein}
\fpage{1}
\centerline{\textbf{INFLUENCE OF THE VAN HOVE SINGULARITY ON}}
\vspace*{0.035truein}
\centerline{\textbf{THE SPECIFIC HEAT JUMP IN BCS SUPERCONDUCTORS}}

\vspace*{0.37truein}
\centerline{
   \footnotesize
     TODOR MISHONOV,$^{\dag,\ddag,}
   $\footnote{Corresponding author; phone: (++32) 16 327183, fax: (++32) 16 327983,\\
              e-mail:~\texttt{todor.mishonov@fys.kuleuven.ac.be}}\\ STEFAN-LUDWIG DRECHSLER$^{\S}$ and
              EVGENI PENEV$^{\ddag}$}
\vspace*{0.015truein}
\centerline{$^{\dag}$\footnotesize\it
   Laboratorium voor Vaste-Stoffysica en Magnetisme, %
   Katholieke Universiteit Leuven
}
\baselineskip=10pt
\centerline{\footnotesize\it
   Celestijnenlaan 200 D, B-3001 Leuven, Belgium
}
\vspace*{0.015truein}
\centerline{$^\ddag$\footnotesize\it
   Department of Theoretical Physics,
   Faculty of Physics, Sofia University ``St. Kliment Ohridski''
}
\baselineskip=10pt
\centerline{\footnotesize\it
    5~J.~Bourchier Blvd., 1164 Sofia, Bulgaria
}
\vspace*{0.015truein}
\centerline{$^{\S}$\footnotesize\it
   Institut f\"ur Festk\"orper- und Werkstofforshung Dresden,
   D-001171 Dresden, Germany
}
\vspace*{0.225truein}

\vspace*{0.21truein}
\abstracts{Within the weak-coupling BCS scheme we derive a general
form of the coefficients in the Ginzburg-Landau expansion of the free
energy of a superconductor for the case of a Fermi level close to a
van Hove singularity (VHS). A simple expression for the influence of
the VHS on the specific heat jump is then obtained for the case where
gaps for different bands are distinct but nearly constant at the
corresponding sheets of the Fermi surface.}{}{}

\vspace*{0.21truein}

\vspace*{1pt}\textlineskip

\section{Introduction}
\label{sec:intro}
\vspace*{-0.5pt}
\noindent

The influence of a van Hove singularity (VHS) on the properties of
superconductors is a largely debated problem in physics of
superconductivity.\cite{Newns:95} Thus, we shall attempt here to
analyze the influence of VHS in the density of states (DOS) on the
jump of the specific heat $\Delta C$ at the critical temperature
\Tc. This problem reduces to finding the coefficients of the Ginzburg-Landau (GL)
expansion of the free energy.
Particular attention will be paid to the account of \emph{multigap} effects.
Let us recall that 44 years ago Moskalenko\cite{Moskalenko:59}
predicted the existence of \textit{multigap} superconductivity, in
which a disparity of the pairing interaction in different bands, such
as the $s$ and $d$ bands in transition metals, leads to different
order parameters and to an enhancement of the critical
temperature. Subsequently multiband effects in superconductors were
intensively investigated, see for example \Rref{Suhl:59,Moskalenko:65,Palistrant:67} and references therein.

Multiband superconductors show small values of $\Delta C(\Tc)$, negative curvature of the upper critical
magnetic field $H_{\mathrm{c2}}(T)$ near the transition temperature, etc.
One of the fundamental properties of multigap superconductors is
that nonmagnetic impurities are pairbreaking;\cite{Palistrant:67}
for a nice introduction in the properties of multiband superconductors
the reader is referred to the review by Moskalenko, Palistrant and
Vakalyuk.\cite{Moskalenko:91} Here we consider the case of clean superconductors.

\section{Model and notation}

To begin with, let us introduce standard notations for the
dimensionless quasimomentum $\bp$ and momentum-space
averaging in the $D$-dimensional case,
\begin{equation}
 \frac{1}{\cal N}
 \sum_{\bp} f(\bp) = \int_0^{2\pi}\!\!\dots\int_0^{2\pi}
 \frac{d\bp}{(2\pi)^D}\, f(\bp) \equiv \la f \ra_{\bp},
\label{eq:mean}
\end{equation}
where ${\cal N}$ is the number of $\bp$-points in the momentum-space
summation or, equivalently, the number of unit cells in the
crystal. It is also expedient to introduce a non-normalized
integration over the Fermi surface
\begin{align}
\la f(\bp) \ra_{\text{F}} & \equiv \la
\delta(\eps_{b,\bp}-E_{\mathrm{F}}) f(\bp) \ra_{\bp}\nn\\
 &=\sum_b
\idotsint
 \delta(\eps_{b,\bp}-E_{\mathrm{F}}) f(\bp) \frac{d\bp}{(2\pi)^D} =
 \sum_b \idotsint\limits_{\eps_{b,\bp} = E_{\mathrm{F}}}f(\bp) \frac{d
 S_{b,\bp}}{v_{b,\bp}(2\pi)^D},
\end{align}
where $dS_{b,\bp}$ is an infinitesimal surface element of the
Fermi surface sheet of the $b$th energy band,
and $\mathbf{v}_{b,\bp}=\nabla_{\bp}\eps_{b,\bp}$ is the quasiparticle
``velocity'' which according to the present convention has dimension
of energy. Conversion to the true velocity in m/sec can be performed
by multiplying with the lattice constant and dividing by $\hbar$. In
this notation for the electronic DOS per unit cell and spin we have
\begin{equation}
\rho(E_\mathrm{F})=\la 1 \ra_{\text{F}}.
\end{equation}
We assume that the DOS can be represented as a sum of a regular around
the Fermi level function $\rho_1$ and a singular part $\rho_2$
divergent at the energy of the van Hove transition $E_{\mathrm{VHS}},$
\begin{equation}
\rho(\eps)=\rho_1(\eps)+\rho_2(\eps), \quad
\begin{cases}
\rho_1(\eps) = \text{const}, & \eps = E_{\mathrm{F}} ,\\
\rho_2(\eps) = \infty, & \eps = E_{\mathrm{VHS}}
\end{cases}.
\end{equation}
Using the so introduced DOS the well-known formula for the normal
specific heat per unit cell reads
\begin{align}
 \frac{C_{\mathrm{n}}}{\cal N}
 & = 2\kb\la\frac{\nu_{\bp}^2}{\cosh^2(\nu_{\bp})}\ra_{\!\!\bp} \nn \\
 & = \frac{2}{3}\pi^2\kb^2T
     \left(\rho_{1}(E_\mathrm{F})
     +\int_{-\infty}^{+\infty}\rho_2(E_\mathrm{F}+2\kb
 T\,\nu)q_c(\nu)d\nu\right),
\label{eq:Cn}
\end{align}
where
\begin{equation}
 \nu_{b,\bp}=\frac{\varepsilon_{b,\bp}-E_\mathrm{F}}{2k_\mathrm{B}T}, \quad
 q_c(\nu)= \frac{6}{\pi^2} \left(\frac{\nu}{\cosh\,\nu}\right)^2, \qquad
 \int_{-\infty}^{+\infty}q_c(\nu)d\nu=1,
\end{equation}
(see also \Fref{fig:1}).

\section{Ginzburg-Landau coefficients}

The starting point of our thermodynamic analysis is the GL expansion
for the free energy per unit cell as a function of the temperature $T$ and
the order parameter $\Xi:$
\begin{equation}
\frac{F(\Xi,T)}{\cal{N}}=a_0\frac{T-\Tc}{\Tc}\,|\Xi|^2+\frac{1}{2}b|\Xi|^4.
\end{equation}
In thermodynamics of second-order phase transitions the ratio of the
GL coefficients $a_0$ and $b$ determines the jump of the specific
heat per unit cell at the critical temperature, so for
superconductors
\begin{equation}
  \frac{1}{\cal N}(C_{\mathrm{s}}- C_{\mathrm{n}})|_{\Tc}
  = \frac{\Delta C}{\cal N} = \frac{1}{\Tc} \frac{a_0^2}{b},
\label{eq:DeltaC}
\end{equation}
where $C_{\mathrm{s}}$ is the specific heat of the superconducting
phase.

Employing the finite-temperature Bogoliubov-Valatin variational
approach\cite{Bogoliubov:57,Abrikosov:58,Abrikosov:88} we have recently given a simple derivation\cite{Mishonov:02}
of Gor'kov and Melik-Barkhudarov's\cite{Gorkov:64} results for the GL coefficients of a clean
weak-coupling superconductor with anisotropic gap,
\begin{equation}
\Delta_{b,\bp}(T)=\Xi(T)\chi_{b,\bp}.
\end{equation}
The gap anisotropy factor $\chi_{b,\bp}$ is the eigenfunction of the
linearized BCS gap equation, corresponding to the maximal
eigenvalue. Performing appropriate modification of these results,
Eqs.~(27, 30) of \Rref{Mishonov:02} can be written in the form
\begin{align}
a_0 =&\; \la \left| \chi_{\bp} \right|^2\ra_{\text{F},1}
   + \frac{1}{4\kb\Tc}
   \la\frac{|\chi_{\bp}|^2}{\cosh^2\nu_{\bp}}\ra_{\!\!\bp,2}\nn\\
b =&\; \frac{7 \zeta(3)}{8\pi^2 (\kb\Tc)^2}\la \left|
  \chi_{\bp} \right|^4\ra_{\text{F},1}
  +\frac{1}{4(2\kb\Tc)^3}\la|\chi_{\bp}|^4\,Q(\nu_{\bp})\ra_{\bp,2},
\end{align}
where
\begin{equation}
Q(\nu) = \frac{1}{\nu^2}\left( \frac{\tanh \nu}
         {\nu} - \frac{1}{\cosh^2\nu}\right).
\label{eq:Q}
\end{equation}
Here we have partitioned the bands into a regular and a singular
one. For the case of regular bands only the general expression for the relative
jump of the specific heat reads\cite{Mishonov:02,Pokrovsky:63}
\begin{equation}
 \left.\frac{\Delta C}{C_{\mathrm{n}}}\right|_{\Tc}
 = \frac{12}{7\zeta(3)} \frac{1}{\beta_{\Delta}}, \qquad
   \frac{1}{\beta_{\Delta}}  = \frac{ \la|\Delta_{\bp}|^2\ra_{\text{F}}^2}{%
             \la 1 \ra_{\text{F}} \la |\Delta_{\bp}|^4 \ra_{\text{F}}
} \leqslant 1,\qquad
\frac{12}{7\zeta(3)} = 1.42613\dots,
\label{eq:jump}
\end{equation}
where the last numerical value is the universal BCS ratio. The question we pose now is
whether the existence of a VHS can change the inequality
(\ref{eq:jump}) and thus drive an enhancement of the relative specific heat jump,
 $1/\beta_{\Delta}>1$. For the latter we shall derive a
suitable for experimental data processing formula.

\section{Influence of the van Hove singularity}

\begin{figure}
\centering
\includegraphics[width=0.7\textwidth]{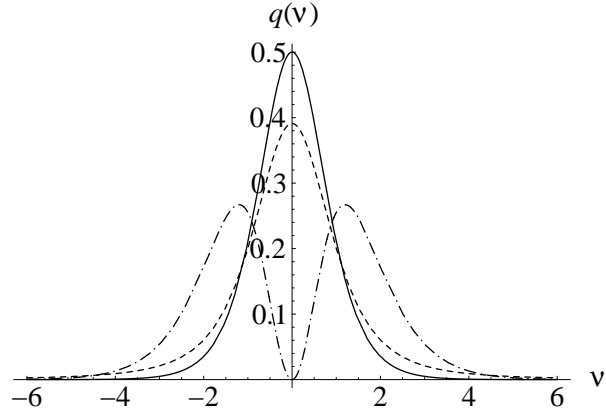}
\caption{Plot of the $q_a$ (solid line), $q_b$ (dashed line), and $q_c$
(dash-dotted line) functions.
\label{fig:1}}
\end{figure}

We assume that the sample is characterized with sufficient purity, and
that the VHS is close to the Fermi level,
\begin{equation}
\hbar/\tau_c \ll \kb \Tc \simeq |E_\mathrm{F}-E_\mathrm{VHS}|,
\end{equation}
where $\tau_c$ is the scattering time. In analogy to the normal
specific heat (\ref{eq:Cn}), for the GL coefficients we can perform the
averaging over the Fermi surface for all regular bands. For the
band having VHS, instead, one has to carry out a separate summation over the
different constant-energy layers in the momentum space:
\begin{align}
 a_0 = & \; \overline{|\chi_1|^2}\,\rho_{1} (E_\mathrm{F}) +
 \overline{|\chi_2|}^2\int_{-\infty}^{+\infty}
 \rho_2(E_\mathrm{F}+2\kb \Tc\,\nu)q_a(\nu)d\nu,\nn\\
 b = &\; \frac{7
 \zeta(3)}{8\pi^2 (\kb\Tc)^2}
 \left[\overline{|\chi_1|^4}\,\rho_{1}(E_\mathrm{F}) +
 \overline{|\chi_2|^4}\int_{-\infty}^{+\infty}
 \rho_2(E_\mathrm{F}+2\kb \Tc \,\nu)q_b(\nu)d\nu \right],
\end{align}
where
\begin{gather}
 \overline{|\chi_1|^k} = \frac{\la |\chi_{1,\bp}|^k
 \ra_{\mathrm{F},1}}{\la 1 \ra_{\mathrm{F},1}},\nn \\
 \overline{|\chi_2|^k} = \lim_{E\rightarrow E_{\mathrm{F}}}
 \int\limits_{\eps_{2,\bp} = E} |\chi_{2,\bp}|^k \frac{d
 S_{2,\bp}}{v_{2,\bp}(2\pi)^D} \left[\int\limits_{\eps_{2,\bp} = E} \frac{d
 S_{2,\bp}}{v_{2,\bp}(2\pi)^D}\right]^{-1},
\end{gather}
with $k=2,\,4,$ and
\begin{equation}
q_a(\nu) = \frac{1}{2}\frac{1}{\cosh^2 \nu},\qquad
q_b(\nu) = \frac{\pi^2}{14\zeta(3)} \left(\frac{\tanh\nu}{\nu} -
           \frac{1}{\cosh^2\nu}\right)\frac{1}{\nu^2}.
\end{equation}
The same normalization as for $q_c$ holds for the $q_{a,b}$ functions:
$\int_{-\infty}^{+\infty}q_{a,b}(\nu)d\nu=1.$ These functions are
shown in \Fref{fig:1}. If, in an acceptable approximation, we can
consider the gaps in the two different bands to be nearly constant,
the sign of the energy-surface averaging can be dropped:
\begin{align}
a_0 = & \; |\chi_1|^2\,\rho_{1} (E_\mathrm{F})
+ |\chi_2|^2\int_{-\infty}^{+\infty} \rho_2(E_\mathrm{F}+2\kb \Tc\,\nu)q_a(\nu)d\nu,\nn\\
b = &\; \frac{7 \zeta(3)}{8\pi^2 (\kb\Tc)^2}
  \left[|\chi_1|^4\,\rho_{1}(E_\mathrm{F})
   + |\chi_2|^4\int_{-\infty}^{+\infty}
   \rho_2(E_\mathrm{F}+2\kb \Tc \,\nu)q_b(\nu)d\nu
  \right].
  \label{eq:coeff2}
\end{align}
The generalization of these expressions for the multiband case is straightforward.
Using \Eref{eq:coeff2}, we find for the specific heat jump
\begin{equation}
 \left.\frac{\Delta C}{C_{\mathrm{n}}}\right|_{\Tc} = \frac{12}{7\zeta(3)}\,
 \frac{1}{\beta_{\mathrm{VHS}}},
\end{equation}
where for the renormalizing multiplier we obtain a generalized
two-band expression, cf. Refs.~\citen{Moskalenko:59,Pokrovsky:63,Mishonov:02}:
\begin{align}
\frac{1}{\beta_{\mathrm{VHS}}}= &
  \left[\overline{|\Delta_1|^2}\,\rho_{1}(E_\mathrm{F}) + \overline{|\Delta_2|^2}\int_{-\infty}^{+\infty}
   \rho_2(\eps_{\nu})q_a(\nu)d\nu\right]^2 \nn \\
 & \times
   \left[\rho_{1}(E_\mathrm{F}) +
   \int_{-\infty}^{+\infty}\rho_2(\eps_{\nu})q_c(\nu)d\nu\right]^{-1} \nn \\
  & \quad\times \left[\overline{|\Delta_1|^4}\,\rho_{1}(E_\mathrm{F})
    + \overline{|\Delta_2|^4}\int_{-\infty}^{+\infty}
    \rho_2(\eps_{\nu}) q_b(\nu) d\nu \right]^{-1},
    \label{Eq:betaDelta}
\end{align}
with
\begin{equation}
 \eps_\nu\equiv E_{\mathrm{F}} + 2 \kb \Tc\,\nu.
\end{equation}
In the multiband case a summation over the band
index is required, and the generalized Moskalenko-Pokrovsky formula reads
\begin{align}
 \left.\frac{\Delta C}{C_{\mathrm{n}}}\right|_{\Tc}
 &= \frac{12}{7\zeta(3)}
  \left[\sum_b \overline{|\Delta_b|^2}\int_{-\infty}^{+\infty}
   \rho_b(\eps_{\nu})q_a(\nu)d\nu\right]^2 \nn \\
 &\qquad\quad \times
   \left[\sum_b
   \int_{-\infty}^{+\infty}\rho_b(\eps_{\nu})q_c(\nu)d\nu\right]^{-1} \nn \\
  &\qquad\quad \quad\times \left[\sum_b
    \overline{|\Delta_b|^4}\int_{-\infty}^{+\infty}
    \rho_b(\eps_{\nu}) q_b(\nu) d\nu \right]^{-1}.
    \label{Eq:betaDeltaMany}
\end{align}
Note that the GL order parameter $\Xi$ in \Eref{Eq:betaDelta} cancels. This
is related to the conditional factorization
\begin{gather}
\Delta_\bp = \Xi \chi_\bp=(c\Xi) (\chi_\bp/c),\nn\\
a_0|\Xi|^2 =(a_0/c^2)|c\Xi|^2 \propto |\chi_\bp|^2 |\Xi|^2
           = (|\chi_\bp|^2/c^2)|c\Xi|^2,\nn\\
b|\Xi|^4 = (b/c^4) |c\Xi|^4 \propto |\chi_\bp|^4 |\Xi|^4
         = (|\chi_\bp|^4/c^4)|c\Xi|^4,\nn\\
\frac{\Delta C}{\cal N} = \frac{1}{\Tc} \frac{a_0^2}{b}
           = \frac{1}{\Tc} \frac{(a_0/c^2)^2}{(b/c^4)},\nn\\
\frac{F(\Xi,T)}{\cal{N}} = a_0\frac{T-\Tc}{\Tc}\,|\Xi|^2+\frac{1}{2}b|\Xi|^4
            =(a_0/c^2)\frac{T-\Tc}{\Tc}\,|c\Xi|^2+\frac{1}{2}(b/c^4)|c\Xi|^4,
\end{gather}
which conserves as scaling invariants all measurable quantities like, e.g.,
the heat capacity and the free energy. Using this freedom we can
normalize $\overline{|\chi_\bp|^2}=1.$

\begin{figure}[t]
\includegraphics[width=0.4\textwidth]{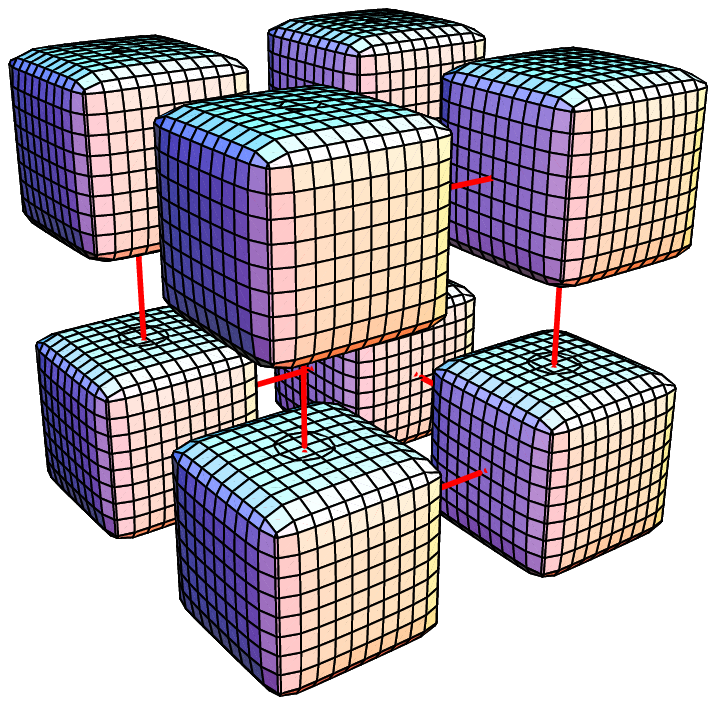}\hfill
\includegraphics[width=0.4\textwidth]{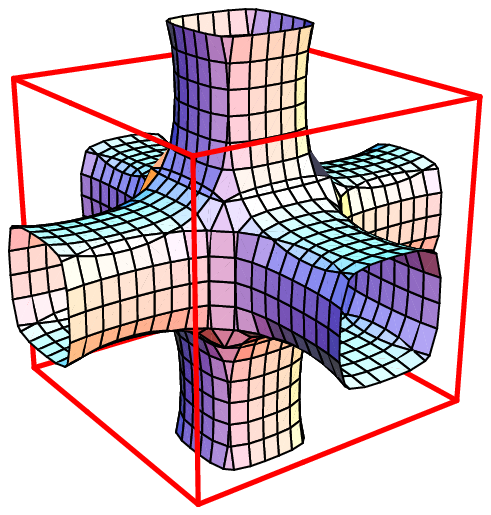}
\caption{Typical constant energy surfaces for cubic perovskites.\protect\cite{Mishonov:97} When the ``dices'' (left)
 become cubes the perfect nesting of their flat surfaces creates 1D singularity ($\propto 1/\sqrt{\varepsilon-E_{\mathrm{VHS}}}$)
of the DOS. Similarly, the nearly constant cross-section of the narrow tubes (right) creates a 2D singularity of the
DOS. The box indicates the first Brillouin zone. \label{fig:2} }
\end{figure}

Analyzing \Eref{Eq:betaDelta} one can easily verify that for the model
cases of one-dimensional (1D) and two-dimensional (2D) VHS, respectively,
\begin{align}
\rho_2^{\mathrm{(1D)}}(\varepsilon)  & = (\eps-E_\mathrm{VHS})^{-1/2},
 \eps - E_{\mathrm{VHS}} > 0\nn\\
 \rho_2^{\mathrm{(2D)}}(\varepsilon) & =  -\ln|\varepsilon-E_\mathrm{VHS}|,
\end{align}
the renormalizing multiplier $\beta_{\mathrm{VHS}}^{-1}$ could be
$>1.$ In this case the VHS can enhance the relative jump $\Delta C/C_{\mathrm{n}}|_{\Tc}$
to values bigger that the conventional $1.43$. For getting an insight into the
influence of the VHS we recommend the DOS $\rho_2(E_{\mathrm{F}} + 2 \kb
\Tc\,\nu)$ to be plotted along with the $q_{a,b,c}(\nu)$ functions. There is no
mathematical restriction how big $1/\beta_{\mathrm{VHS}}$ could be for
a very narrow peek of the DOS. However, realistic DOS corresponds to 1D or 2D case.
Such type of singularities appear for cubic perovskites\cite{Mishonov:97} where
close to the VHS some constant-energy surfaces have the shape of elongated
tubes (2D cross section), or dices (1D cross section). Some typical
surfaces are shown in \Fref{fig:2}. Just the opposite situation could
arise, however, in a two-band VHS model if the $E_{\mathrm{F}}$ lies in between
 two logarithmic VHS. In this case $C_{\mathrm{n}}$ can be
considerably enhanced. We find it instructive for a realistic
VHS model the thermodynamic variables, e.g.,
$\Delta C/C_{\mathrm{n}}(\Tc)$, $\Delta C$ and $C_{\mathrm{n}}(\Tc),$ to be
plotted versus $E_{\mathrm{F}}-E_{\mathrm{VHS}}.$ The $C_{\mathrm{n}}(T)/T$ dependence
for some particular $(E_\mathrm{F}-E_\mathrm{VHS})$-values is of general interest even far from
\Tc.

Concluding, we believe that the influence of the VHS can be studied by focusing on a very
simple characteristic---the jump of the specific heat at $T_c$.
Such measurements do not require large single crystals.

\nonumsection{Acknowledgments}
\noindent
The authors gratefully acknowledge the stimulating discussions with
Joseph Indekeu, his hospitality and interest in this study.
One of the authors (T.~M.) is thankful to A.~A.~Abrikosov, J.~Bouvier, L.~P.~Gor'kov, V.~A.~Moskalenko,
D.~M.~Newns, M.~E.~Palistrant, V.~L.~Pokrovsky, and C.~C.~Tsuei for
the correspondence related to their papers.
This work was partially supported by GOA and V.~Mishonova.

%
\nonumsection{References}


\begin{thebibliography}{99}

\bibitem{Newns:95} D.~M.~Newns, C.~C.~Tsuei and P.~C.~Pattnaik,
 \jour{\PR}{52}{13611}{1995};
 C.~C.~Tsuei, C.~C.~Chi, D.~M.~Newns, P.~C.~Pattnaik, and D\"aumling,
 \jour{\PRL}{69}{2134}{1992};
 J.~Friedel, \jour{J. Phys.: Condens. Matt.}{1}{7757}{1989};
 J.~Labbe and J.~Bok, \jour{Europhys. Lett.}{3}{1225}{1987};
 J.~Bouvier and J.~Bok, \jour{J. Superconductivity}{10}{673}{1997};
 J.~Bouvier and J.~Bok,
 \jour{Physica}{C364-365}{471}{2001};
 J. Bouvier and J. Bok, \jour{Physica}{C288}{217}{1997};
 R.~S.~Markiewicz, \jour{J. Phys.: Condens. Matt.}{2}{665}{1990};
 R.~S.~Markiewicz, \jour{J. Phys. Chem. Solids}{58}{1179--1310}{1997},
 and references there in, Appendix A;
 Z.~Szotek, B.~L.~Giorffy, W.~M.~Temmerman, and O.~K.~Andersen,
 \jour{\PR}{B58}{522}{1998};
 P.~A.~Lee and N.~Reed, \jour{Phys. Rev. Lett.}{58}{2691}{1987};
 A.~A.~Abrikosov, \jour{\PR}{B56}{446}{1997}.

\bibitem{Moskalenko:59} V.~A.~Moskalenko,
\jour{Fiz. Met. i Metalloved.}{8}{503}{1959}
[\jour{Phys. Met. Metallogr. (USSR)}{8}{25}{1959}];
Due to priority reasons we have to stress that this paper bares a receive date ``31 October 1958'', and was
published in October 1959 before the submital of any other paper on multigap superconductivity. We feel
emphaty to the discrimination of this classical paper by the present MgB$_2$-community,
especially by those familiar with the history of the two-band model.

\bibitem{Suhl:59} H.~Suhl, B.~T.~Mattias, and L.~R.~Walker,
\jour{\PRL}{3}{552}{1959}.

\bibitem{Moskalenko:65} V.~A.~Moskalenko and M.~E.~Palistrant,
\jour{Zh.~Exp.~Teor.~Fiz.}{49}{770}{1965}
[\jour{Sov. Phys. JETP}{22}{536}{1966}];
%
V.~A.~Moskalenko, \textit{On the theory of solid state},
Doctoral Dissertation,
(Moskow, 1966) (in Russian) (unpublished).
%
V.~A.~Moskalenko and M.~E.~Palistrant,
\jour{Dokl. Akad. Nauk. SSSR}{162}{532}{1965}
[\jour{Sov. Phys. Dokl.}{10}{457}{1965}]
%
V.~A.~Moskalenko,
\jour{Zh.~Exp.~Teor.~Fiz.}{51}{1163}{1966}
[\jour{Sov. Phys. JETP}{24}{780}{1966}];
%
V.~A.~Moskalenko,
\jour{Fiz. Met. i Metalloved.}{23}{585}{1967}
[\jour{Phys. Met. Metallorg. (USSR)}{}{9}{1967}];
%
V.~A.~Moskalenko and M.~E.~Palistrant,
``\textit{Theory of pure two-band superconductors}'' in
\textit{Statistical Physics and Quantum Field Theory -- in memoriam to
S.~V.~Tyablikov} (in Russian: \textit{Staitisticheskaya fizika i
kvantovaya teoriya polya}), ed. N.~N.~Bogoliubov (Moskow, Nauka,
1973), p.~226;
%
V.~A.~Moskalenko, \textit{Method of investigation of
density of states of superconducting alloys} (Kishinev, Stiinta, 1974)
(in Russian);
%
V.~A.~Moskalenko, \textit{Electromagnetic and kinetic
properties of superconducting alloys with overlapping bands}
(Kishinev, Stiinta, 1976) (in Russian);
%
V.~A.~Moskalenko, Yu.~N.~Nica and D.~F.~Digor,
\textit{Tunneling properties of superconducting alloys}
(Kishinev, Stiinta, 1978) (in Russian);
%
V.~A.~Moskalenko, M.~E.~Palistrant and L.~Z.~Kon,
\textit{Low temperature properties of metals with singularities of energy bands},
(Kishinev, Stiinta, 1989) (in Russian);
%
V.~A.~Moskalenko, M.~E.~Palistrant, and V.~M.~Vakalyuk
\jour{Fiz. Nizk. Temp}{15}{378}{1989}
[\jour{Sov. J. Low Temp. Phys.}{15}{213}{1989}];
%
V.~A.~Moskalenko, M.~E.~Palistrant, V.~M.~Vakalyuk, and I.~V.~Padure,
\jour{Solid State Commun.}{69}{747}{1989};
%
V.~A.~Moskalenko, M.~E.~Palistrant, V.~M.~Vakalyuk,
\textit{Xth Int. Symp. on Jahn-Teller Effect},
(Kishinev, Stiinta 1989) p.~88.

\bibitem{Palistrant:67} M.~E.~Palistrant and V.~I.~Dedju,
\jour{Phys. Lett.}{A24}{537}{1967};
%
L.~Z.~Kon,
\jour{Fiz. Met. i Metalloved.}{23}{211}{1967}
[\jour{Phys. Met. Metallogr. (USSR)}{23}{17}{1967}];
%
M.~K.~Kolpajiu, L.~Z.~Kon.
\jour{Izvestia AN MSSR. Seria fiz. mat. nauk.}{12}{90}{1967}
(Proceedings of Moldavian Academy of Science, In Russian);
%
M.~E.~Palistrant and M.~K.~Kolpadjiu,
\jour{Phys. Lett.}{A41}{123}{1972};
%
M.~E.~Palistrant and M.~K.~Kolpadjiu,
\textit{Quantum theory of many-particle systems},
(Kishinev, Stiinta, 1973) (in Russian);
%
M.~E.~Palistrant, A.~T.~Trifan, and K.~K.~Gudima
\jour{Fiz. Nizk. Temp.}{}{452}{1976}
[\jour{Sov. J. Low.Temp. Phys.}{2}{224}{1976}];
%
M.~E.~Palistrant and A.~T.~Trifan,
\jour{Fiz. Nizk. Temp.}{3}{976}{1977}
[\jour{Sov. J. Low. Temp. Phys.}{3}{473}{1977}]
%
M.~E.~Palistrant and A.~T.~Trifan,
\textit{Theory of doped superconductors under pressure},
(Kishinev, Stiinta, 1980) (in Russian);
%
M.~E.~Palistrant and F.G.~Kochorbe,
\jour{Physica}{C194}{351--356}{1992};
%
F.~G.~Kochorbe and M.~E.~Palistrant,
\jour{Zh. Eksp. Teor. Fiz.}{77}{442}{1993};
%
F.~G.~Kochorbe and M.~E.~Palistrant,
\jour{Theor. Math. Phys.}{96}{1083}{1993};
%
M.~E.~Palistrant,
\jour{J. Superconductivity}{10}{19}{1997};
%
F.~G.~Kochorbe and M.~E.~Palistrant,
\jour{Physica}{C298}{217}{1997};
%
M.~E.~Palistrant, F.~G.~Kochorbe,
\jour{J. Low Temp. Phys.}{26}{299}{2000}.

\bibitem{Moskalenko:91}
V.~A.~Moskalenko, M.~E.~Palistrant, and V.~M.~Vakalyuk,
\jour{Usp.~Fiz.~Nauk}{161}{155}{1991}
[\jour{Sov.~Phys.~Usp.}{34}{717}{1991}].

\bibitem{Bogoliubov:57} N.~N.~Bogoliubov, D.~N.~Zubarev and Yu.~A.~Tserkovnikov,
\jour{Dokl. Acad. Nauk SSSR}{177}{788}{1957} [\jour{Sov. Phys. Dokl.}{2}{535}{1958}];
%
N.~N.~Bogoliubov,
\jour{Zh.~Exp.~Teor.~Fiz.}{34}{58}{1958} [\jour{Sov. Phys. JETP}{7}{41}{1957}];
%
N.~N.~Bogoliubov,
\jour{Nuovo Cimento}{7}{794}{1958};
%
N.~N.~Bogoliubov, N.~N.~Tolmachev and D.~V.~Schirkov,
\textit{A new method in the theory of superconductivity} (Moscow, Nauka, 1958) (in Russian),
English translation (New York, Consultants Bureau, 1958);
%
J.~G.~Valatin, \jour{Nuovo Cimento}{7}{843}{1958}; J.~G.~Valatin and D.~Butler, \jour{Nuovo
Cimento}{10}{37}{1958}.

\bibitem{Mishonov:02} T.~Mishonov and E.~Penev,
\jour{Int. J. Mod. Phys.}{B16}{}{2002} (in print), cond-mat/0206118.

\bibitem{Abrikosov:58} A.~A.~Abrikosov and I.~M.~Khalatnikov,
\jour{Uspekhi Fiz. Nauk.}{95}{551}{1958}
[\jour{Sov. Uspekhi}{}{}{}].

\bibitem{Abrikosov:88} A.~A.~Abrikosov, \textit{Fundamentals of the
Theory of Metals} (North Holland, Amsterdam, 1988)
Sec.~16.4 and Sec.~17.1 and references therein.
%

\bibitem{Gorkov:64} L.~P.~Gor'kov and T.~K.~Melik-Barkhudarov,
\jour{Zh.~Exp.~Teor.~Fiz.}{45}{1493}{1963}
[\jour{Sov. Phys. JETP}{18}{1031}{1964}].

\bibitem{Pokrovsky:63} V.~L.~Pokrovski\u\i\ and M.~S.~Ryvkin,
\jour{Zh.~Exp.~Teor.~Fiz.}{43}{92}{1962}
[\jour{Sov. Phys. JETP}{16}{67}{1963}]; V.~L.~Pokrovskii,
\jour{Zh.~Exp.~Teor.~Fiz.}{40}{641}{1961}
[\jour{Sov. Phys. JETP}{13}{447}{1961}].
%
\bibitem{Mishonov:97} T.~M.~Mishonov, I.~N.~Gentchev, R.~K.~Koleva and
E. S. Penev, \jour{Superlatt. Miscrostruct.}{21}{471}{1997};
T.~M.~Mishonov, I.~N.~Genchev, R.~K.~Koleva and E. S. Penev,
\jour{Superlatt. Miscrostruct.}{21}{477}{1997}.

\end{thebibliography}
\end{document}